\documentclass[pre,twocolumn]{revtex4-1}%
\usepackage{amsmath}
\usepackage{graphicx}
\usepackage[latin9]{inputenc}
\usepackage[T1]{fontenc}
\usepackage[english]{babel}
\usepackage{amsfonts}
\usepackage{amssymb}

\newcommand{\be}{\begin{equation}}
\newcommand{\ee}{\end{equation}}
\newcommand{\ben}{\begin{eqnarray}}
\newcommand{\een}{\end{eqnarray}}

\newcommand{\corr}{{\ensuremath{\mathrm{Corr}}}}
\newcommand{\cov}{{\ensuremath{\mathrm{Cov}}}}
\newcommand{\dist}{{\ensuremath{\mathrm{dist}}}}

\begin{document}
\title{Space-time correlations in urban sprawl}
\author{A. Hernando}
\email{alberto.hernandodecastro@epfl.ch}
\affiliation{Laboratory of Theoretical Physical Chemistry, Institut des Sciences et
Ing\'enierie Chimiques, \'Ecole Polytechnique F\'ed\'erale de Lausanne,
CH-1015 Lausanne, Switzerland}
\author{R. Hernando}
\affiliation{Social Thermodynamics Applied Research (SThAR), Madrid, Spain}
\author{A. Plastino}
\affiliation{National University La Plata, Physics Institute (IFLP-CCT-CONICET)  La Plata, Argentina}
\affiliation{Universitat de les Illes Balears and IFISC-CSIC, Palma de Mallorca, Spain}

\begin{abstract}
Understanding demographic and migrational patterns constitutes a great
challenge. Millions of individual decisions, motivated by
economic, political, demographic, rational, and/or emotional
reasons underlie the high complexity of demographic dynamics.
Significant advances in quantitatively understanding such
complexity  have been registered in recent years,
as those involving the growth of cities
[Bettencourt LMA, Lobo J, Helbing D, Kuehnert C,  West GB (2007) Growth,. Innovation, Scaling, and the Pace of Life in Cities, 
{\it Proc Natl Acad Sci USA} 104 (17) 7301-7306]
but many fundamental issues still defy comprehension.
We present here compelling empirical evidence of a high level of
regularity regarding time and spatial correlations in urban sprawl,
unraveling patterns about the {\it inertia} in the growth of cities and
their {\it interaction} with each other.
By using one of the
world's most exhaustive extant demographic data basis
---that of the Spanish Government's Institute INE,
with records covering  111 years and (in 2011) 45 million people,
distributed amongst more than 8000 population nuclei---
we show that the inertia of city growth has a characteristic
time of 15 years, and its interaction with the growth of other cities
has a characteristic distance of 70 km. Distance is shown to be the main
factor that entangles two
cities (a 60\% of total correlations). We present a mathematical
model for population flows that i) reproduces  all these
regularities and ii) can be used to predict the
population-evolution of cities. The  power of our current social
theories is thereby enhanced.
\end{abstract}
\date{\today}
\maketitle

\section{Introduction}

The quantitative description of social human patterns is one of
the great challenges of this century. Significant advances have
been achieved in understanding the complexity of city growth,
urban sprawl, electoral elections, and many other social systems
 \cite{1sta,Natu,PnasW,Pnas2,bat1,city1,city2,power,zipf,ciudad2,firms,net2,elec1,gibrat,oppi,mob1,mob2}.
One finds that the concomitant patterns can be successfully
modelled, involving subjacent universal scaling properties
\cite{PnasW,Pnas2,nosEPJB10,nosPLA12} and fundamental principles
---as the Maximum Entropy \cite{X1,nosEPJB12,nosEPJB12b,nosJRSI13,nosPRE12}
or the Minimum Fisher Information \cite{nosPLA09,nosPA10} ones.
Also, the interaction between cities (as measured by, for instance, the
number of crossed phone calls\cite{gmodel} or  human
mobility\cite{mob1}) displays  predictable characteristics. Thus,
it is plausible to conjecture that some kind of universality
underlies collective human behavior\cite{nosPRE12,mob2}.

However, many fundamental issues still defy comprehension. Our aim
in this work is to answer two question regarding city growth and
human migrations: i) is the growth of cities inertial? i.e., does
the population growth in the present year depend on the growth of
past years? and ii) does the growth of a city depend on the growth
of  neighboring cities? i.e., does the migration of people from
one city to other exhibit spatial patterns? Millions of individual
decisions, motivated by economic, political, demographic,
rational, and/or emotional reasons underlies the growth rate of a
city. Accordingly, one may  expect some level of randomness and
unpredictability. In this vein, one might think that \newline
 i) if some inertia is present, the growth rate of the present year could be
deduced from that in  past years, and \newline
ii) if some correlation with other cities exists, the growth rate might be
predicted from the rates of  other cities.\\
Thus, the observation and detection of regular space-time patterns in
urban-population evolution could be viewed as constituting an
important step towards understanding collective, human dynamics at
the macro-scale. Indeed, the parameterization of such regularities
could lead
to a potential improvement of the present population-projection tools 
and analysis \cite{pnas3,tools}.\\

\subsection{Urban growth} 
The evolution of city population has been
described with great success in the past by recourse to
geometrical Brownian walkers obeying a dynamical equation that
exhibits scale-invariance
\cite{city1,city2,gibrat,nosPLA12,nosEPJB12b,nosJRSI13,nosPRE12}
\begin{equation} 
\dot{X}_i(t) = v_i(t) X_i(t),
\end{equation}
where $X_i(t)$ is the population at time $t$ of the $i$-th city (of an ensemble of $n$ cities),
$\dot{X}_i(t)$ stands for  its temporal change, and $v_i(t)$ for
the growth-rate. One finds in the literature that this rate
usually displays stochastic behavior in the form of a Wiener
process that complies with $\langle
v_i(t)v_j(t')\rangle=\sigma_v^2\delta_{ij}\delta(t-t')$, so that
we deal with  uncorrelated noise. In spite of its simplicity, this
reductionist model is able to describe many of the observations
reported for city-rank distributions. Indeed, this equation can be
linearized by defining $u_i(t)=\log[X_i(t)]$ thus obtaining
\begin{equation}  
\dot{u}_i(t) = v_i(t),
\end{equation}
which allows one to recover all well-known properties of regular
Brownian motion   \cite{nosEPJB12b}. Indeed, a ``thermodynamics of
urban population flows"
---with the pertinent observables--- can be derived following the
analogy with physics presented in Ref. \cite{nosPRE12}.
However, uncorrelated evolution is assumed in \cite{nosPRE12} for
the sake of simplicity, which entails operating with the
equivalent of a scale-free ideal gas. Such an  assumption was
sufficient for explaining  the main properties of the macroscopic
state of an ensemble of cities, but a higher-level theory that
would provide deeper understanding is desirable. Indeed, some sort
of {\it interaction} between cities is of course to be expected,
as well as some kind of inertia. The ensuing correlations are of
great importance to understand the complex patters of migration
and to improve our  predictive power with regards to the subjacent
dynamics.

\section{Results}

An exhaustive census data-set is indeed needed, something not
easy to come by. Fortunately, the Spanish Government's Institute
INE \cite{ine} provides information about the population of
8100 municipalities ---the smallest administrative unit--- during
111 years, from 1900 to 2011. They are distributed on a surface of
$\sim500,000$ km$^2$ inhabited by  more than 45 million people
(2011). Fig. 1a displays the spatial distribution of the Spanish
municipalities, and Fig. 1b their time-evolution. A typical
diffusion pattern is visible. The population's median and
arithmetic mean are also plotted. The former has grown with  time
but the later has diminished, telling us that the population has
descended in a majority of towns, which reflects on the migration
from country-side to large cities, a common pattern in most of the
world. The diffusion process is readily discernible: one
appreciates that the width of the distribution does grow.

\subsection{Statistical properties of growth rates} 
In order to analyze in more detail the underlying dynamics, we
base our considerations on the developments of Refs.~
\cite{nosJRSI13,nosEPJB12b,nosPRE12}. It is shown there that  the
dynamical growth equation for city populations exhibits  the
general appearance
\begin{equation}
\dot{X}_i(t) = v_i(t)X_i(t) + w_i(t)\sqrt{X_i(t)}, \label{2terms}
\end{equation}
where $w_i(t)$ is a Wiener coefficient. We face proportional
growth in the first term to which  a finite-size contribution
(FSC) is added in the second one. The later becomes small for
large sizes but is important for small ones. The second term can
be regarded as 'noise' and is thus expected to be independent of
the proportional growth. Accordingly, the variance $V[\dot{X}_i]$
can be written as
\begin{equation}\label{var}
V[\dot{X}_i]/X_i = \sigma_{vi}^2 X_i + \sigma_{wi}^2,
\end{equation}
where $\sigma_{vi}$ and $\sigma_{wi}$ are the associated deviations
of $v_i$ and $w_i$, respectively.

Comparison with the data entails appealing to numerical time
derivatives for each $\dot{X}_i$. We use yearly data from 1996
till 2011 (whenever the appropriate data-sets are available for
each intermediate year) so as to generate the graph of Fig.~1c,
that displays the $(X_i,V[\dot{X}_i]/X_i)-$pairs for all the
Spanish municipalities computed as
\begin{eqnarray}\label{means}
\langle\dot{X}_i\rangle &=& \displaystyle\frac{1}{T}\sum_{t=1}^{T}\dot{X}_i(t),\\
\displaystyle
V[\dot{X}_i]&=&\langle[\dot{X}_i-\langle\dot{X}_i\rangle]^2\rangle\nonumber\\
&=&\frac{1}{T}\sum_{t=1}^{T}(\dot{X}_i(t)-\langle\dot{X}_i\rangle)^2.
\end{eqnarray}
where $T=14$ is the total number of data-sets used for this
calculation. The median $\mathrm{med}(V[\dot{X}_i]/X_i)$ nicely
fits Eq.~(\ref{var}), with $\sigma_{v} = 0.0119$ and
$\sigma_{w}=0.47$, respectively. Notice that FSC fluctuations are
larger than multiplicative ones, the later dominating, of course,
for large sizes. The transition between both regimes occurs at
$x_T=\sigma_{wi}^2/\sigma_{vi}^2=1500$ inhabitants.\\

\begin{figure}[ht]
\begin{center}
\centerline{\includegraphics[width=\linewidth,trim= 0 0 0 220,clip]{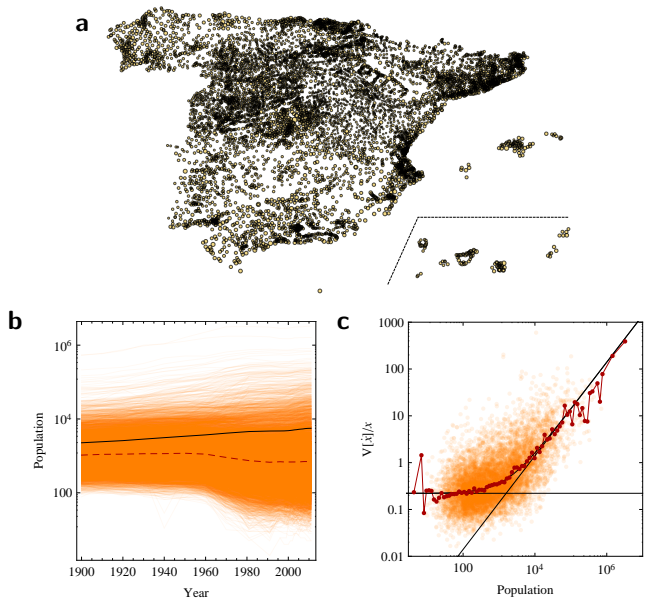}}
\caption{Characteristics of the dataset.
{\bf a,} Spatial distribution of Spanish
Municipalities. Circle's sizes are proportional to the
population's logarithm. {\bf b,}  Evolution of  all
municipalities. We give also the population per town (arithmetic
mean in black and geometric mean in red). {\bf c,}
Variance of the  population-change  vs. population for each
municipality (dots). The median value (red dot-line) clearly
follows Eq. (\ref{var}) (black line).}\label{fig1}
\end{center}
\end{figure}

\subsection{Empirical observation of inertial growth} 
To find whether
there exists a systematic dependence between successive yearly
growths (or inertia) we consider first the $n$-cities-average and
variance such that
\begin{eqnarray}\label{sumt}
 \langle \dot{x}(t)\rangle &=&
\frac{1}{n}\sum_{i=1}^n \dot{x}_i(t),\\
V[\dot{x}(t)]
&=&
\frac{1}{n}\sum_{i=1}^n[\dot{x}_i(t)-\langle\dot{x}(t)\rangle]^2,
\end{eqnarray}
where $x_i(t)=X_i(t)/N(t)$ with $N(t)$ the total population at
time $t$, excluding in this fashion the effects of the total
population growth. Time correlations have been obtained via the
Pearson product-moment correlation coefficient ($\corr$) between
data-sets pertaining to different years $t_1$ and $t_2$. The mean
correlation as a function of the time interval $\Delta t=|t_1-t_2|$ is obtained as the average 
\begin{eqnarray}\label{cdt} 
c(\Delta t)&=&\frac{1}{T}\sum_{t=1}^{T}\mathrm{Corr}[\dot{x}(t),\dot{x}(t+\Delta t)]\nonumber\\
&=&\frac{1}{T}\sum_{t=1}^{T}\frac{\mathrm{Cov}[\dot{x}(t),\dot{x}(t+\Delta t)]}{\sqrt{V[\dot{x}(t)]V[\dot{x}(t+\Delta t)]}},
\end{eqnarray} 
where $\cov(a,b)$ is the covariance between variables $a\,-\,b$ and $T$
is now the total number of available data-sets for each case. We study first
such correlations as a function of the population window, where
two different situations are encountered. Within a standard
deviation, no correlations exist for low populations, but they are
significative for large ones, as indicated by Fig.~2a. The
transition between the two ensuing regimes takes place at
populations of $\sim 1000$ inhabitants. Thus, for the finite size
term in (\ref{var}) no time correlations are detected. They do
appear, though, in the proportional growth regime. Accordingly, we
evaluate time-correlations for municipalities with populations of
more that ten thousand inhabitants  during a period of up to 50
years. We find that correlations decay as the time interval
$\Delta t=|t_1-t_2|$ between observations increases (Fig.~2b). The
resulting mean value can be nicely fitted by an exponential
function
\begin{equation}\label{eqex}
\langle\corr(\Delta t)\rangle  = c_t\exp(-\Delta t/\tau),
\end{equation}
with $c_t = 0.74\pm0.02$ and $\tau = 15\pm1$ years. {\it Accordingly,
the correlation's mean-time in the demographic flux is of around
15 years.}\\

\subsection{Empirical observation of spatial correlations.} 
We pass now
to a study of the demographical entanglement between two given
cities, as represented by spatial correlations. The correlation
coefficient between the $i$-th and $j$-th city reads
\begin{equation}\label{cij}
c_{ij}=\mathrm{Corr}[\dot{x}_i,\dot{x}_j]=\frac{\mathrm{Cov}[\dot{x}_i,\dot{x}_j]}
{\sqrt{V[\dot{x}_i]V[\dot{x}_j]}}, 
\end{equation} 
where the covariances,
variances, and means are time-averages as in Eq.~(\ref{means}).
Amongst a host of possible entanglement factors, we choose here to
study  the simplest one: distance between cities $\Delta r$.
Accordingly, we evaluate correlations between  cities versus their
 pertinent distance $\dist(i,j)$ via the histogram 
\begin{equation}
\langle\mathrm{Corr}(\Delta r)\rangle=\frac{1}{n}\sum_{i=1}^n
c_{ij}\delta(\Delta r-\dist(i,j)). 
\end{equation} 
We find that for towns with
more that 10000 inhabitants --within the proportional growth
regime-- the mean value of the spatial correlation does depend
upon distance as a power law, but saturates for short distances.
Things can be nicely fitted by the expression
\begin{equation} \label{lorenzo}  
\langle\corr(\Delta r)\rangle = \frac{c_r}{1+|\Delta r/r_0|^\alpha}, 
\end{equation}
obtaining $c_r=0.33\pm0.02$, $r_0=76\pm10$, and
$\alpha=1.8\pm0.3$, with a coefficient of determination $R^2$
equal to $0.9159$. Instead, fixing for future convenience
$\alpha=2$, that yields a Lorentz function, we get
$c_r=0.33\pm0.01$ and $r_0=79\pm8$, with $R^2=0.9156$. Since the
concomitant  two ways of fitting are indistinguishable, we adopt
$\alpha=2$ for simplicity. {\it As a consequence, the
typical ``demographic distance" turns out to be (in average) of
$\sim 80$~km, decaying with $r^{-2}$ at large distances.} Thus, we
face long-range correlations (Fig. 2d). The influences of {\it
other} factors, though, make these correlations to vanish at about
500 km. We use our data to compare i) the width of $\corr(\Delta
r)$ with ii) that expected for a bivariate normal distribution
\cite{biv} (see Appendix). The empiric width is larger than
the bivariate one: $0.327$ vs. $0.204$  (Fig. 2c), indicative of the
presence of additional, distance-independent, correlations. We
deduce that the separation
 between towns, that is, their mutual distance, is the origin of about a 60\% of the total
correlation between them.\\

\begin{figure}[ht]
\begin{center}
\centerline{\includegraphics[width=\linewidth,trim= 0 0 0 220,clip]{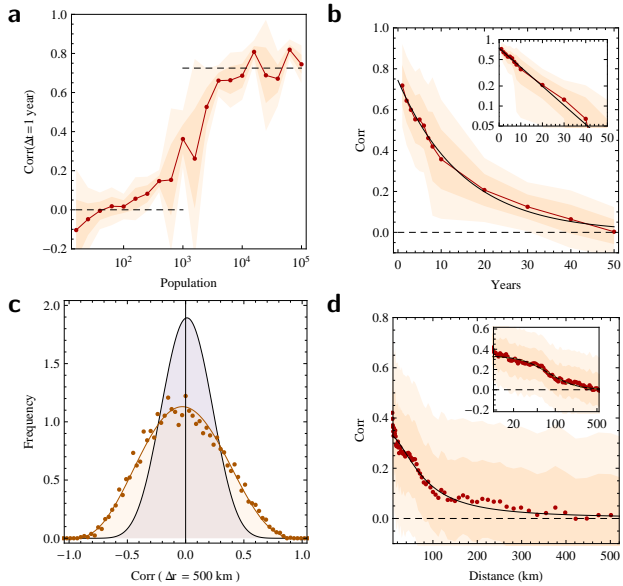}}
\caption{Empirical space-time correlations of population growth. 
{\bf a,} Time-correlations versus town-sizes for
yearly relative growths (red curve and red dots). The shaded area
represents the width determined by one standard deviation (darker
hue) and by two of them (lighter hue). Horizontal dark lines are
just visual aids. {\bf b,} Time correlation for the
relative growth of towns populated by more that 10,000
inhabitants. Shaded areas represent  widths determined by one
(darker hue) and two (lighter hue) standard deviations,
respectively. Inset: same representation, but in a log scale. The
exponential fit of Eq.~(\ref{eqex}) acquires thus more visibility.
{\bf c,} Comparison of widths: bivariate-normal vs.
empirical
 correlations-distribution.
{\bf d,} Spatial correlation of Spain's municipalities'
relative growth for populations larger than  10,000 inhabitants.
In black, the Lorentz shape of Eq.~(\ref{lorenzo}) (for
$\alpha=2$) compared with the empirical mean (red dots). Inset:
same representation, with a log scale for the distance.}\label{fig2}
\end{center}
\end{figure}

\subsection{Quantitative model for inertial and correlated urban growth}
How to explain and reproduce these remarkable
results? To such an end we advance  here a model,
compatible with previous descriptions and observations,  inspired
by the Langevin equation    \cite{langevin}. Accordingly, it
includes inertia, `forces' $F_i(t)$, and a friction-coefficient
$\gamma$, whose values should fit empirical observation.
Correlated forces imply a correlation matrix $Q_{ii'} = \langle
F_i(t) F_i'(t)\rangle/V_f$, where $V_f$ is the variance of the
forces, to be empirically adjusted. Disregarding finite-size noise
one is led to
\begin{eqnarray}
F_i(t)       &=& \sum_{i'=1}^n R_{ii'}f_{i'}(t),\label{w1}\\
\dot{v}_i(t) &=& F_i(t) - \gamma v_i(t),\label{w2}\\
\dot{u}_i(t) &=& v_i(t),\label{w3}\\
\dot{x}_i(t) &=& e^{u_i(t)},\label{w4}
\end{eqnarray}
where 
\begin{itemize} 
\item $f_{i}(t)$ are uncorrelated random
forces such that $\corr[f_{i}(t),f_{i'}(t')]=V_f\delta_{ii'}\delta(t-t')$ and \item
$R_{ii'}$ are the matrix elements of a correlation-generating
matrix such that $\sum_{i'j'}R_{ii'}R_{jj'}=Q_{ij}$.
\end{itemize} 
The form of $F_i(t)$ suggests that the force acting on
a city is somewhat the average value of several independent ones.
Now, an important personal decision is that of
selecting to move to a certain location on the basis of available
information. This information derives from human contacts of the
concomitant individual, whose spatial distribution (SD) has been
found to follow a $r^{-2}$ law at large distances, saturating for
short ones~\cite{gmodel}. For simplicity, we assume a Lorentz
shape for this SD
\begin{equation}
R_{ij} = \frac{R_j(0)}{1+|2\Delta r_{ij}/r_0|^2},
\end{equation}
where $\Delta r_{ij}$ is the distance between the $i$ and $j$-th cities
and the normalization constant is defined as
\begin{equation}
R_j(0) =  \left[\sum_{k=1}^n(1+|2\Delta r_{kj}/r_0|^2)^{-2}\right]^{-\frac{1}{2}}.
\end{equation}
Thus, $F$ becomes a ``coarse-grained" force.
Let us consider for our derivation of $Q$ the continuous limit
$x_i\rightarrow x(\mathbf{r})$, with $\mathbf{r}$ a planar spatial
coordinate. $x(\mathbf{r})$ represents the relative population at
$\mathbf{r}$, and the total normalized population becomes
$1=\int_Sd\mathbf{r}x(\mathbf{r}),$ where  $S$ is the pertinent
region's area. Since we deal now with the coordinates $\mathbf{r}$
and $\mathbf{r}'$ instead of the indexes $i,j$, the $R$
matrix-elements are a function $R(|\mathbf{r}-\mathbf{r}'|)$.
Sums become integrals obtaining $R(0)=2[2/\pi r_0^2]^{1/4}$ 
and the convolution ($\otimes$) for the coarse-grained force
\begin{equation}
F(\mathbf{r},t) = R(\mathbf{r})\otimes f(\mathbf{r},t) =
\int_Sd\mathbf{r}'\frac{2[2/\pi r_0^2]^{1/4}f(\mathbf{r}',t)}{1+4|(\mathbf{r}-\mathbf{r}')/r_0|^2}.
\end{equation} 
Since the convolution of two Lorentzians of equal scale is
also a Lorentzian with twice that scale-parameter, we find for the
forces-correlation
\begin{eqnarray} 
\mathrm{Corr}[F(\mathbf{r}),F(\mathbf{r}')]&=&Q(|\mathbf{r}-\mathbf{r}'|)\nonumber\\
&=&R(|\mathbf{r}-\mathbf{r}'|)\otimes R(|\mathbf{r}-\mathbf{r}'|)\nonumber\\
&=&\frac{1}{1+|(\mathbf{r}-\mathbf{r}')/r_0|^2}.
\end{eqnarray}
Thus we write for the general case
\begin{equation}\label{q}
Q(\Delta r_{ij}) = \frac{1}{1+|\Delta r_{ij}/r_0|^2}.
\end{equation}
To obtain the correlation for the growth we solve Eqs. (\ref{w1})-(\ref{w4}) for $v$ and $x$ writing
\begin{eqnarray}
v_i(t) &=& e^{-\gamma t}v_i(0) + \int_0^td\tau e^{-\gamma(t-\tau)}F_i(\tau).\\
x_i(t) &=& \exp\left[\frac{v_i(0)}{\gamma}(1-e^{-\gamma t})+\right.\nonumber\\
      &&\phantom{\exp~~}\left.\int_0^td\tau\int_0^\tau d\tau' e^{-\gamma(\tau-\tau')}F_i(\tau')\right].
\end{eqnarray}
We have then $C_{ij}(\Delta t)=\mathrm{Corr}[\dot{x}_i(t),\dot{x}_j(t+\Delta t)]=\mathrm{Corr}[v_i(t),v_j(t+\Delta t)]$.
On the basis of that the initial time is arbitrary, we assume $t\rightarrow\infty$
so as to obtain the $v-$correlation
\begin{equation}\label{esso}
C_{ij}(\Delta t)=\mathrm{Corr}[v_i(t),v_j(t+\Delta t)] =\frac{e^{-\gamma\Delta t}}{1+|\Delta r_{ij}/r_0|^2}
\end{equation}
which nicely reproduces empirical data with $\gamma=1/\tau$
(from the variance of $v_i(t)$ we also obtain $V_f/2\gamma=\sigma_v^2$).

Without trying to be exhaustive, we have tested our equations
with a numerical experiment. One simulates a square (area) of
500$\times$500 km$^2$, and randomly  place on it  1000 ``virtual"
cities (Fig.~3a). Using the empirical values for $r_0$, $\gamma$,
and $V_f$, one makes the system to  evolve during  100 years. All
cities possess the same population at the beginning. The
concomitant results are analyzed  by recourse to the methods used
above for dealing with empirical data. Comparisons are made with
theoretical predictions and  plotted in Fig.~3b and 3c for time
and spatial correlations, respectively. Expectations are seen to
be  fulfilled. It is worth mentioning that we have followed a
normal-modes description to solve the associated equations,
working with collective, independent modes (see Appendix). Our virtual
municipalities display the same behavior recorded for actual ones.
The main difference ensues from the presence of (as yet) undefined
correlations in the empirical data.

\begin{figure}[ht]
\begin{center}
\centerline{\includegraphics[width=\linewidth,trim= 0 0 0 220,clip]{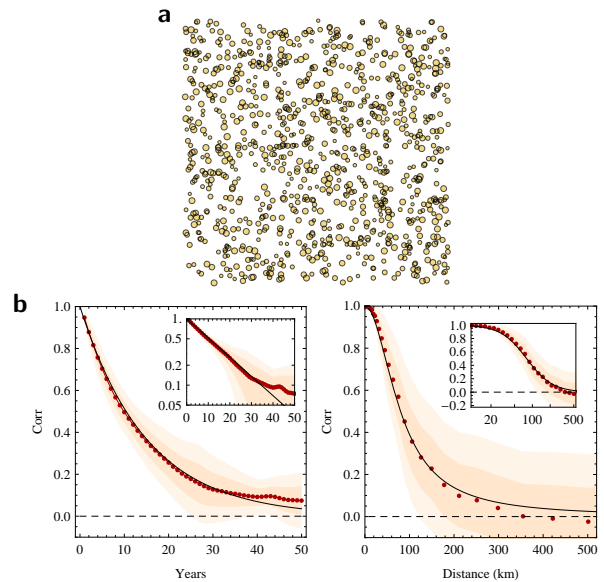}}
\caption{Numerical experiment solving Eqs.(\ref{w1})-(\ref{w4}).  
{\bf a,} Spatial distribution of  ``virtual"
municipalities. Each side of the square represents 500~km. {\bf
b,} Time correlation for the relative growth of virtual towns.
Shaded areas represent  widths determined by one (darker hue) and
two (lighter hue) standard deviations, respectively. Inset: same
plot in a log scale. {\bf c,} Spatial correlation of virtual
municipalities' relative growth. In black, the theoretical Lorentz
shape of Eq.~(\ref{q}), compared with the empirical mean (red
dots).}\label{fig3}
\end{center}
\end{figure}

\section{Conclussion}

Summing up, by recourse to the geometric walkers-model of
Eqs.~(\ref{w1}-\ref{w4}), we have empirically demonstrated that the relative
growth of a city's  population  exhibits both i) inertia and ii)
correlation with the relative growth of neighboring cities, with
distance as the main variable that underlies that town-town
interaction. We also showed that a model inspired by the Langevin
equation is able to reproduce these observations. 
Indeed, the model that we present here can be used to improve the predictive 
power of present techniques for demographic projection. However, further
improvements are needed in order to identify the {\it  undefined
correlations within the actual data} whose existence we have
discovered.  We expect that these correlations will depend on
local circumstances and also on the particular socio-economic
status of each city.\\

{\it Acknowledgments.}  This work was  partially supported by Social Thermodynamics 
Applied Research (SThAR) (to AH and RH), and the  project PIP1177 of
CONICET (Argentina), and the projects FIS2008-00781/FIS
(MICINN)-FEDER, EU, Spain (to AR).

\appendix
\section{Distribution of correlation coefficients}

For a bivariate normal distribution, the distribution of correlation coefficients is given by
\begin{eqnarray}
P(c,C,T) &=& \frac{1}{\sqrt{2\pi}}(T-2)\frac{\Gamma(T-1)}{\Gamma(T-1/2)}(1-c^2)^{T/2-2}\times\nonumber\\
&&\left[1-C^2\right]^{(T-1)/2}\left[1-Cc\right]^{T-3/2}\times\nonumber\\
&&{}_2F_1\left[1/2,1/2,T-1/2,(Cc+1)/2\right],\label{pcc} 
\end{eqnarray}
where $c$ stands for the correlation-value that one might
numerically obtain using Eq. (\ref{cij}), $C$ is the actual
correlation value and $T$ the number of data-point used 
to evaluate $c$.

\begin{figure*}[ht]
\begin{center}
\centerline{\includegraphics[width=\linewidth,trim= 0 0 0 0,clip]{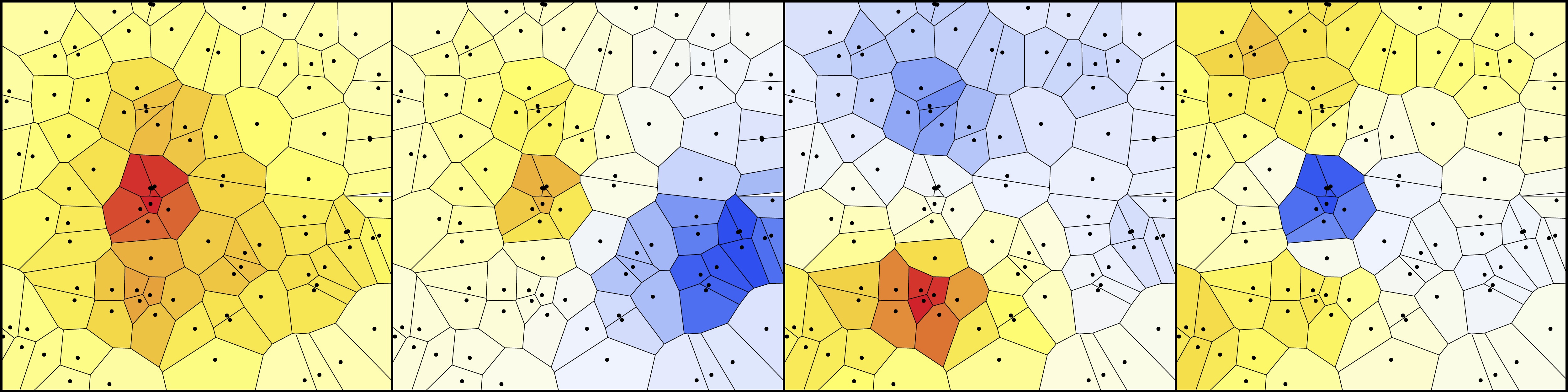}}
\caption{omponents of the first four eigenvectos of the simulated system (from white to red: positive
values; from white to blue: negative values). The surface of each municipality is defined by its Voronoi area.}\label{fig4}
\end{center}
\end{figure*}

\section{The normal mode solution for the correlated Langevin equation}

The computational cost of solving Eqs. (\ref{w1})-(\ref{w4}) can be reduced via
 a normal-mode treatment. Indeed, we have defined a change-of-basis matrix  $A$ such that
$R$ (and $Q$) become diagonal. This generates new variables
$u'_i(t)=\sum_{i'}A_{ii'}\log[x_i(t)]$ whose motion-equations are
\begin{eqnarray}
\dot{u}'_i(t) &=& v'_i(t),\\
\dot{v}'_i(t) &=& \sqrt{\varepsilon_i}f'_i(t) - \gamma v'_i(t),
\end{eqnarray}
with  $v'_i(t)=\sum_{ii'}A_{ii'}v_i(t)$,
$f'_i(t)=\sum_{ii'}A_{ii'}f_i'(t)$, and $\sqrt{\varepsilon_i}$ is
the $i$-th eigenvalue of $R$ (with  $\varepsilon_i$ that of $Q$).
One easily checks that the forces $f'$ are statistically
equivalent to those indicated by $f$ (i.e.,
$\langle f'_i(t)f'_j(t+\Delta t)\rangle = V_f\delta_{ij}\delta(\Delta t)$
), so that the simulation involves directly
the random generation of $f'$, without having to actually effect the
basis-change. The variables $u'_i(t)$ evolve in independent
fashion, representing normal-mode evolution. The presence of
$\sqrt{\varepsilon_i}$ accounts for different mode-equilibrations
between  $f'$ and the dumping $\gamma$. This fact might be
conceived as originating mass-factors. Figure~\ref{fig4} displays the first
four modes for 100 cities distributed uniformly in a square of 100$\times$100~km
using $r_0=30$~km, in such a way that the color at the Municipality  $i$
represents the coefficient $A_{ii'}$ for the eigenvector $i'=$1, 2, 3 and 4
(the surface of each virtual municipality is in this example the Voronoi area).

\end{document}